\documentclass{webofc}
\usepackage[varg]{txfonts}   
\usepackage{amsmath}
\usepackage{bm}

\begin{document}
\title{Nonequilibrium evolution of quarkonium in medium}
%
%

\author{\firstname{Yukinao} \lastname{Akamatsu}\inst{1}\fnsep\thanks{\email{akamatsu@kern.phys.sci.osaka-u.ac.jp}}
\and \firstname{Takahiro} \lastname{Miura}\inst{1}
}

\institute{Department of Physics, Osaka University, Toyonaka, Osaka 560-0043, Japan}

\abstract{%
We review recent progress in open quantum system approach to the description of quarkonium in the quark-gluon plasma.
A particular emphasis is put on the Lindblad equations for quarkonium and its numerical simulations.
}
\maketitle
\section{Introduction}
\label{intro}
Relativistic heavy-ion collision experiment provides us with a unique opportunity to study the novel matter composed of quarks and gluons, which are permanently confined inside hadrons at the energy scales of our daily experiences.
This novel matter, which is called Quark-Gluon Plasma (QGP), exhibits hydrodynamic collective behavior during the expansion process.
The mechanism of early thermalization and the applicability of hydrodynamics to smaller size systems remain fundamental questions in the heavy-ion physics.

In the heavy-ion collisions, a quarkonium acts as an impurity for the QGP and can probe its local color-electric fluctuations\footnote{
In this article, the word ``quarkonium'' is used to mean a quantum mechanical state of heavy quark and antiquark, which is not necessarily a bound state or an eigenstate.
} \cite{Rothkopf:2019ipj}.
Single heavy quark can also probe the same property of the QGP, but there is a big difference: a quarkonium is made of heavy quark-antiquark pair interacting with each other through attractive force if they are in the color singlet.
This attractive force binds the heavy quark pair so that its color is hidden inside the localized wave function.
In this sense, a quarkonium is a sensitive probe of QGP at higher temperatures, typically determined by its bound state size.
During the last decade, there have been a lot of developments in understanding the non-equilibrium evolution of quarkonium in the QGP as an open quantum system \cite{Akamatsu:2020ypb, Yao:2021lus, Sharma:2021vvu}, which we summarize in this article. 

Even though the complexity of heavy-ion collisions is enormous, the first step to describe a quarkonium as an  open system is to derive a master equation in the simplest setup, namely the environment is a static and homogeneous QGP and the heavy quarks are the non-relativistic quantum mechanical particles.
This simplification implicitly assumes that the (pseudo)critical temperature $T_c$ is much lower than the heavy quark mass $M$ so that one can find a temperature $T_c < T \ll M$. 
The effects which are yet to be included are as follows:
\begin{itemize}
\item Interaction between initially uncorrelated heavy quark pairs.\\
-- This could be justified for bottomonium because it is rare that two bottom quark pairs are produced simultaneously in a heavy-ion collision.
For charmonium, its justification is subtle at RHIC energy scales and probably not given at the LHC.
For this reason, our description better applies to bottomonium.
\item The effects of non-static and inhomogeneous QGP evolution.\\
-- The master equation contains parameters determined by the temperature, which is assumed to be a constant.
Choosing the temperature by the local temperature at (the quantum expectation value of) the center of mass for quarkonium, the above effects can be partially taken into account.
\item Heavy quark pair creation in the QGP.\\
-- This is justified because such process is suppressed by a factor $e^{-2M/T}\ll 1$ with $2M\gg T$.
\item Heavy quark pair annihilation in the QGP.\\
-- This process is suppressed by $1/M^2$ and can be added perturbatively if needed.
\end{itemize}

The second step is to numerically solve the master equation.
Practically, the center-of-mass motion of quarkonium is assumed to follow a free-streaming classical trajectory determined by some model for its initial phase space distribution, and the master equation is solved only for the relative motion.
As is the case for the matter evolution, the evolution of quarkonium during the medium thermalization process is not precisely known at present.
Therefore, simulation of the master equation starts at the medium thermalization time $\tau_{\rm therm.}$, which can be estimated to be $\sim \mathcal (0.4\text{-}1) \ {\rm fm}$ by the initial time of the hydrodynamic simulations at RHIC \cite{Hirano:2002ds} and the LHC \cite{Song:2010mg}.
To perform the simulation, we need to determine the initial wave function and the freezeout description of quarkonium.

If we ignore the complication from the influences of thermalizing quarks and gluons and color gauge fields, the formation time of quarkonium bound state is estimated by the inverse of its binding energy $\tau_{\rm form.}\sim 1/E_{\rm bind.}$, which is $\sim (1\text{-}2) \ {\rm fm}$ for the bottomonium ground state in the Coulomb potential.
Then, the initial wave function of quarkonium at $\tau_{\rm therm.}\lesssim \tau_{\rm form.}$ cannot be too different from that at the production, which is approximated by a delta function, up to possible corrections from thermalizing process involving saturation scales.
Note that if binding energy is measured from the $B\bar B$ threshold, we get $\tau_{\rm form.}\sim 0.2 \ {\rm fm}$ and it is more appropriate to evaluate $\tau_{\rm therm.}\sim \tau_{\rm form.}$.
In any case, the initial wave function of quarkonium at $\tau_{\rm therm.}$ has not yet been fully investigated and it is customary to choose a singlet or octet Gaussian wave packet (as a proxy for the delta function) or a singlet bound state wave function.
Similarly to the initial wave function, the freezeout description of quarkonium at later stage of the heavy-ion collisions is not very much known.
As an ad-hoc description, a freezeout temperature $T_f\sim T_c$ is chosen at which we stop the simulation and project the density matrix onto a vacuum state in the singlet.

With the overall picture given as above, we will now review two essential ingredients in the open system descriptions for quarkonium --- the master equation (Sec.~\ref{sec:master_equation}) and its numerical simulation (Sec.~\ref{sec:numerical_simulation}).

\section{Lindblad equations for quarkonium}
\label{sec:master_equation}
\subsection{Basics of open quantum systems}
\label{sec:basics}
When a small physical system of our interest is surrounded by environmental degrees of freedom, we can obtain an effective description by integrating out the latter.
To be more specific, the total system is given on a Hilbert space $\mathcal H_{\rm tot} = \mathcal H_S\otimes \mathcal H_E$ with a Hamiltonian $ H_{\rm tot} = H_S\otimes I_E + I_S\otimes H_E + H_I$, and the information of the system quantum states is contained in the reduced density matrix $\rho_S\equiv {\rm Tr}_E\rho_{\rm tot}$.
The master equation is an evolution equation of $\rho_S$.
It was found about 45 years ago that if the master equation is Markovian and describes a trace-preserving and completely positive evolution of $\rho_S$, it must be written in the Lindblad form \cite{Gorini:1975nb, Lindblad:1975ef}:
\begin{align}
\label{eq:Lindblad}
\frac{d}{dt}\rho_S(t) =\mathcal L(\rho_S) &=-i\left[H_S',\rho_S\right]
+\sum_k \left(L_k\rho_S L_k^{\dagger} - \frac{1}{2}L_k^{\dagger}L_k\rho_S - \frac{1}{2}\rho_S L_k^{\dagger}L_k\right) \\
&=-i\left(H_{\rm eff}\rho_S - \rho_SH_{\rm eff}^{\dagger}\right)
+ \sum_k L_k\rho_S L_k^{\dagger}, \quad
H_{\rm eff} = H_S'  - \frac{i}{2}\sum_k L_k^{\dagger}L_k.
\end{align}
Note that the Hamiltonian $H_S'$ in the Lindblad equation is different from the system Hamiltonian $H_S$ due to the coupling with environment.
One can also rewrite the Lindblad equation as a sum of non-unitary evolution governed by $H_{\rm eff}\neq H_{\rm eff}^{\dagger}$ and the transition or scattering part $\sum_k L_k\rho_S L_k^{\dagger}$.
The in-medium self-energy is given by $H_{\rm eff} - H_S$.

When the system-environment coupling is weak, we can obtain the Lindblad equation based on perturbative expansions.
Following the standard procedure known as Born-Markov approximation \cite{BRE02}, we obtain an evolution equation with similar structure to the Lindblad equation:
\begin{align}
\frac{d}{dt}\rho_S(t) &= \int_0^{\infty} ds \left\langle V_E(s)V_E(0)\right\rangle_T
\left[\begin{aligned}
&V_S(t-s)\rho_S(t) V_S(t) \\
&- V_S(t)  V_S(t-s)\rho_S(t)
\end{aligned}\right] + h.c.  + \mathcal O(V^3),
\end{align}
where $\left\langle V_E(s)V_E(0)\right\rangle_T$ is an environment correlation function in thermal equilibrium.
Here the interaction Hamiltonian is $H_I = V_S\otimes V_E$ and the interaction picture is adopted for the time evolution.
This form already looks similar to the Lindblad equation, but $s$-integration must be performed by approximating $V_S(t-s)$ terms.
In the regime of quantum Brownian motion \cite{Caldeira:1982iu}, it is assumed that the system time scale $\tau_S$ is long compared to the environmental correlation time $\tau_E (\ll \tau_S)$, in which $\left\langle V_E(s)V_E(0)\right\rangle$ changes substantially\footnote{
To derive the pre-master equation in the Born-Markov approximation, one needs to assume that $\tau_E$ is much smaller than the dynamical time scale $\tau_R (\gg \tau_E)$ of $\rho_S(t)$.
The integration range of $s$ can then be extended from $\int_0^t ds$ (if the initial time is $0$) to $\int_0^{\infty}ds$ and the evolution is made Markovian, i.e. dependence on the initial time is lost.
}.
Then, one can approximate as $V_S(t-s) = V_S(t) -is [H_S, V_S(t)] + \cdots$ and obtain after integrating $s$ the Lindblad operator
\begin{align}
\label{eq:Lindblad_op_qbm}
L\propto V_S + \frac{i}{4T} \dot V_S + \mathcal O(V_S^2, \partial_t^2),
\end{align}
using the Kubo-Martin-Schwinger (KMS) relations for $\int_{-\infty}^{\infty}e^{i\omega s}\left\langle V_E(s)V_E(0)\right\rangle_T$.
Note that the derivative expansion in $L$ yields not only the first but also the second order expansion in the master equation, but the precision is ensured only at the first order.

There is a technical remark on \eqref{eq:Lindblad_op_qbm} here.
It must be slightly confusing, but this is the leading order result in the limit $\tau_E/\tau_S\ll 1$ \cite{Akamatsu:2020ypb}.
First, note that the ratio of the first two terms in \eqref{eq:Lindblad_op_qbm} is $\dot V_S/TV_S \sim 1/T\tau_S \ll 1$, but this ratio follows from the KMS relation and does not reflect the smallness of $\tau_E/\tau_S$.
Furthermore, the second term $i\dot V_S/4T$ is an anti-Hermitian correction to the first Hermitian term $V_S$ and can change the dynamical behavior qualitatively because the effective coupling to the environment is now non-Hermitian\footnote{
In the (Lindbladized) Caldeira-Leggett model \cite{diosi1993high, diosi1993calderia, gao1997dissipative, vacchini2000completely}, $L=x+\frac{ip}{4mT}$ and $L\rho L^{\dagger} \sim x\rho x$ part describes decoherence and the cross term $L\rho L^{\dagger} \sim -i(x\rho p - p\rho x)$ describes dissipation.
}.
Therefore, it is not a simple matter to compare them and estimate their relative importance.
So, having both of them can be regarded as a requirement from the fluctuation-dissipation relation of the environment.
The true statement about \eqref{eq:Lindblad_op_qbm} is that this is the Lindblad operator in the leading order in the weak coupling and the leading evaluation in $\tau_E/\tau_S\ll 1$.
There are corrections of order $\tau_E/\tau_S$ to both Hermitian and anti-Hermitian parts of $L$, which may come from the structure of $\int_0^{\infty}e^{i\omega s}\left\langle V_E(s)V_E(0)\right\rangle_T$ at finite $\omega$ and from higher order derivative expansion.

In the application of open quantum system to quarkonium in the QGP, the validity of the quantum Brownian regime is checked as follows.
First, the system time scale $\tau_S$ is the intrinsic time scale of quarkonium and is typically given by the inverse of the bound state energy in the Coulomb potential $V(r) = - C_F\alpha_s/r$.
For the bottomonium ground state, it is
\begin{align}
\tau_S\sim \frac{4}{M(C_F\alpha_s)^2} \sim (1\text{-}2) \ {\rm fm}
\end{align}
for $C_F\alpha_s\sim 0.3\text{-}0.4$ and $M\sim 4.8 \ {\rm GeV}$, which we also called $\tau_{\rm form.}$ in the previous section.
Second, the correlation time of the QGP $\tau_E$ can be estimated by the inverse of the lowest nonzero Matsubara frequency
\begin{align}
\tau_E\sim \frac{1}{2\pi T} \sim \frac{0.16 \ {\rm fm}}{[T/ (0.2 \ {\rm GeV})]}
\end{align}
for a strongly coupled QGP with $T/T_c \sim \mathcal O(1)$.
Then, the hierarchy of the scales $\tau_S\gg\tau_E$ is more or less satisfied.
Note however that the hierarchy $\tau_S\gg \tau_E$ for bottomonium becomes marginal without the factor $2\pi$ in $\tau_E$ and the applicability of quantum Brownian method, i.e. the derivative expansion, is subtle.
Instead, there is a different approach to derive and solve a classical kinetic theory describing the singlet quarkonia bound states as molecules by interpreting $\tau_S\lesssim \tau_E$ \cite{Yao:2017fuc, Yao:2018nmy, Yao:2018sgn, Yao:2020xzw}.

\subsection{Quarkonium described by non-relativistic QCD (NRQCD)}
When the heavy quark mass is larger than any other scales of QCD and the environment such as $\Lambda_{\rm QCD}$ and the temperature $T$, non-relativistic description is available in the heavy quark sector.
One of such non-relativistic effective theories for quarkonium is Non-Relativistic QCD (NRQCD).
In NRQCD, the heavy quark degrees of freedom are the Pauli spinors for heavy quark $\psi$ and antiquark $\chi$ and the effective Lagrangian is written as
\begin{align}
\mathcal L_{\rm NRQCD}
=\mathcal L_{\rm QCD} 
+\psi^{\dagger} \left[iD_t + \frac{\vec D^2}{2M} \right]\psi
+\chi^{\dagger} \left[iD_t - \frac{\vec D^2}{2M} \right]\chi
+\cdots,
\end{align}
at the leading order of the velocity expansion.
For the details of power counting and matching to QCD, see \cite{brambilla2005effective}.
After neglecting the magnetic interaction, the corresponding Hamiltonian is
\begin{align}
H_{\rm tot} &= H_{\rm QCD} + \frac{p_Q^2}{2M} + \frac{p_{Q_c}^2}{2M}
+  gA_0^a(\vec x_Q)t^a_Q - gA_0^a(\vec x_{Q_c})t^{a*}_{Q_c} \\
&= I_S \otimes H_{\rm QCD}
+ \left(\frac{p_Q^2}{2M} + \frac{p_{Q_c}^2}{2M}\right)\otimes I_E
+ \int \frac{d^3 k}{(2\pi)^3} \left[
e^{i\vec k\cdot\vec x_Q} t^a_Q - e^{i\vec k\cdot\vec x_{Q_c}} t^{a*}_{Q_c}
\right] \otimes g\tilde A_0^a(\vec k),
\end{align}
where the second line makes it explicit that the heavy quark part is the system and the light sector constitutes the environment.
Substituting the formula for the Lindblad operators, we get \cite{Akamatsu:2014qsa}
\begin{align}
\label{eq:Lindblad_NRQCD}
\frac{d}{dt}\rho_S(t) &= -i\left[H_S', \rho_S\right]
+\int \frac{d^3 k}{(2\pi)^3} L^a(\vec k)\rho_S L^{a\dagger}(\vec k)
-\frac{1}{2}\left\{
L^{a\dagger}(\vec k)L^a(\vec k), \rho_S
\right\}  + \mathcal O(g^2),\\
L^{a}(\vec k) &= \sqrt{\tilde D(k)}\left[e^{i\vec k\cdot\vec x_Q}t^a_Q - e^{i\vec k\cdot\vec x_{Q_c}}t^{a*}_{Q_c} + \mathcal O(\partial_t)\right],\\
H_S' &=\frac{p_Q^2}{2M} + \frac{p_{Q_c}^2}{2M} +V(r)t^a_Q t^{a*}_{Q_c} + O(\partial_t),
\end{align}
where the real functions $V(r)$ and $D(r)\equiv \int d^3r e^{i\vec k\cdot\vec r}\tilde D(k)$ are defined by the gluon two-point function
\begin{align}
V(r) + iD(r) &=\frac{2i}{N_c^2-1}
\int_0^{\infty}dt \langle gA_0^a(t,\vec r)gA_0^a(0,\vec 0)\rangle_T.
\end{align}
At the length scale of $r\sim 1/gT$, the two-point function receives medium effects from Debye screening mass and Landau damping, which must be resummed to yield the full leading order result:
\begin{align}
V(r) = -\frac{g^2}{4\pi}e^{-m_D r}, \quad
D(r) = g^2T\int\frac{d^3k}{(2\pi)^3}\frac{\pi m_D^2 e^{i\vec k\cdot\vec r}}{k(k^2+m_D^2)^2}.
\end{align}
The in-medium self-energy can be obtained from the Lindblad equation as
\begin{align}
H_S' - \frac{i}{2}\int\frac{d^3k}{(2\pi)^3}L^{a\dagger}(\vec k)L^a(\vec k) -H_S,
\end{align}
whose singlet projection at the leading order in the derivative expansion is the so-called complex potential \cite{Laine:2006ns, Beraudo:2007ky, Brambilla:2008cx}
\begin{align}
V_{\rm complex} (r) =  C_F[V(r) + i(D(r) - D(0))].
\end{align}
For the further details, such as the next-to-leading terms in the derivative expansion and the Lindblad operators in the projected color singlet and octet basis, see \cite{Akamatsu:2014qsa, Akamatsu:2020ypb}.
Also, the master equation at the first order derivative expansion in the strict sense (see Sec.~\ref{sec:basics}) is obtained in \cite{Blaizot:2015hya, Blaizot:2017ypk, Blaizot:2018oev}.

The derivation here relies on the smallness of $g$ and the condition for the quantum Brownian motion reads
\begin{align}
\left.\begin{aligned}
&\tau_E \sim \frac{1}{gT} \quad &(\text{color electric scale}) \\
&\tau_S \sim \frac{4}{M(C_F\alpha_s)^2} \quad &(\text{Coulomb ground state})
\end{aligned}\right\} \qquad
\tau_E \ll \tau_S \Leftrightarrow \frac{M}{T} \ll \frac{4g}{(C_F\alpha_s)^2}.
\end{align}
The master equation can describe phenomena whose dynamical time scale satisfies the Markovian condition $\tau_R \gg \tau_S$ (see the footnote 2).
For example, in order to describe heavy quark kinetic equilibration,
\begin{align}
\tau_R\sim \frac{M}{g^4T^2} \quad (\text{equilibration}) \qquad
\tau_E\ll\tau_R \ \Leftrightarrow \ g^3 \ll \frac{M}{T},
\end{align}
and to describe the decoherence of wave function with size $r\sim 1/m_D$ is
\begin{align}
\tau_R\sim \frac{1}{D(1/m_D)}\sim \frac{1}{g^2T} \quad (\text{decoherence at $r \sim 1/m_D$})\qquad
\tau_E\ll\tau_R \ \Leftrightarrow  \ g \ll 1.
\end{align}

At modestly high temperature $T/T_c\sim \mathcal O(1)$, QGP is expected to be strongly coupled.
In this case, the Lindblad equation derived here is not applicable.
Instead, we model the quantum Brownian motion of quarkonia using the same Lindblad equation, but now with model functions
\begin{align}
\label{eq:model}
C_F V(r) = -\frac{0.3}{r}e^{-2Tr}, \quad
C_F D(r) = \frac{T}{\pi} e^{-(Tr)^2},
\end{align}
which is used later in the simulation discussed in Sec.~\ref{sec:numerical_simulation}.

\subsection{Quarkonium described by potential non-relativistic QCD (pNRQCD)}
When the heavy quark pair is close to each other, there is another effective description, which describes the pair as a color dipole.
The effective field theory is called potential Non-Relativistic QCD (pNRQCD), whose quarkonium degrees of freedom are the singlet $S(t, \vec R,\vec r)$ and the octet $O^a(t, \vec R,\vec r)$ fields.
The fields depend on the center of mass ($\vec R$) and the relative ($\vec r$) coordinates of the quarkonium.
The effective Lagrangian at the leading order in $1/M$ and the next-to-leading order in $r$ is
\begin{align}
\mathcal L_{\rm pNRQCD} &= \mathcal L_{\rm QCD} + \int d^3 r
{\rm Tr}\left[
{\rm S}^{\dagger}\left(i\partial_t - V_s(r) +\cdots\right){\rm S} + {\rm O}^{\dagger}\left(iD_t - V_o(r) +\cdots\right){\rm O}
\right]  \\
&\quad +V_A(r){\rm Tr}\left[
{\rm O}^{\dagger}\vec r\cdot g\vec E {\rm S} + {\rm S}^{\dagger}\vec r\cdot g\vec E {\rm O}
\right]
+\frac{V_B(r)}{2}{\rm Tr}\left[
{\rm O}^{\dagger}\vec r\cdot g\vec E {\rm O} + {\rm O}^{\dagger}{\rm O}\vec r\cdot g\vec E
\right]
+ \cdots, \nonumber \\
{\rm S}(t, \vec R,\vec r)&\equiv \frac{S(t, \vec R,\vec r)}{\sqrt{N_c}}\bm 1, \quad
{\rm O}(t, \vec R,\vec r)\equiv \sqrt{2} O^a(t, \vec R,\vec r)t_F^a,
\end{align}
where $V_s(r) = - C_F\alpha_s/r$, $V_o(r) = \alpha_s/2N_c r$, $V_A(r)=V_B(r)=1$ are the Wilson coefficients at the leading (non-vanishing) order of $\alpha_s$.
Note that the fields in $\mathcal L_{\rm QCD}$ depend on $\vec R$.
Again, for the details of power counting and matching to NRQCD, see \cite{brambilla2005effective}.
By redefining the octet field ${\rm O}(t,\vec R,\vec r)$ and the color electric field $\vec E(t,\vec R)$ so that the covariant derivative $D_t$ becomes $\partial_t$ while keeping the dipole interaction terms unchanged, the theory becomes simpler and the corresponding Hamiltonian is
\begin{align}
H&=H_{\rm QCD} + \frac{p^2}{M} - \frac{C_F\alpha_s}{r}|s\rangle\langle s|
+ \frac{\alpha_s}{2N_c r} |a\rangle\langle a| \nonumber \\
&\quad - \left(
\sqrt{\frac{1}{2N_c}}\left(|a\rangle\langle s| + |s\rangle\langle a|\right) + \frac{1}{2}d^{abc}|b\rangle\langle c|
\right)\vec r\cdot g\vec E^a(\vec R),
\end{align}
where $|s\rangle$ and $|a\rangle \ (a=1,2,\cdots, N_c^2-1)$ are color singlet and octet states for a quarkonium.
Here, the kinetic term $p^2/M$ for the relative motion is added, which balances with the singlet potential in the bound states.
The Lindblad equation in this case reads \cite{Brambilla:2016wgg, Brambilla:2017zei}
\begin{align}
\label{eq:Lindblad_pNRQCD}
\frac{d}{dt}\rho_S(t) &=-i\left[H_S',\rho_S\right]
+ L_{ai}\rho_S L_{ai}^{\dagger} - \frac{1}{2}\left\{L_{ai}^{\dagger}L_{ai}, \rho_S\right\} + \mathcal O(r^3),\\
L_{ai} &= \sqrt{\frac{\kappa}{C_F}}\left(
\sqrt{\frac{1}{2N_c}}\left(|a\rangle\langle s| + |s\rangle\langle a|\right) + \frac{1}{2}d^{abc}|b\rangle\langle c|
\right)\left[r_i + \mathcal O(\partial_t)\right],\\
H_S' &=\frac{p^2}{M} - \frac{C_F\alpha_s}{r}|s\rangle\langle s|
+ \frac{\alpha_s}{2N_c r} |a\rangle\langle a|
+\frac{\gamma}{2C_F}r^2 \left(
C_F|s\rangle\langle s| + \frac{N_c^2-2}{4N_c}|a\rangle\langle a|
\right) + \mathcal O(\partial_t),
\end{align}
where the real coefficients $\gamma$ and $\kappa$ are defined by the two-point function of the color electric field:
\begin{align}
\frac{-\gamma + i\kappa}{C_F} &=\frac{2i}{3(N_c^2-1)}
\int_0^{\infty}dt \langle gE_i^a(t,\vec R)gE_i^a(0,\vec R)\rangle_T.
\end{align}
Note that the two-point function is actually gauge invariant when expressed by the original color electric field.
In this derivation, the small-$r$ expansion for the Lindblad equation does not rely on the smallness of the coupling constant $g$ so that the coefficients $\gamma$ and $\kappa$ are defined non-perturbatively.
The in-medium self-energy for the singlet is $\frac{1}{2}(\gamma - i\kappa)r^2$ at the leading order in the derivative expansion.
The coefficient $\gamma$ causes in-medium mass shift of quarkonium and $\kappa$ gives in-medium width to the quarkonium spectrum as well as the rate of heavy quark momentum diffusion.
Currently available lattice simulations evaluate $\gamma \sim - (0.7\text{-}3.8)T^3$ and $\kappa \sim (0.24\text{-}4.2)T^3$ \cite{Brambilla:2019tpt}.
For the further details, such as the next-to-leading terms in the derivative expansion and the Lindblad operators in the projected color singlet and octet basis, see \cite{Brambilla:2016wgg, Brambilla:2017zei, Akamatsu:2020ypb}.

\section{Simulation of the Lindblad equations for quarkonium} 
\label{sec:numerical_simulation}

Having derived the Lindblad equations, the next step is to solve them.
There is a challenge here: the density matrix has doubled dimensions of the wave function.
So far, several groups have performed numerical simulations of the Lindblad equation for quarkonium in the QGP.
Most of them use a class of simulation methods called stochastic unravelling, in which wave functions are evolved stochastically in such a way that their mixed state density matrix is equivalent to the solution of the Lindblad equation.
To be explicit, with a solution of the stochastic Schr\"odinger equation $\psi_i(\vec x,t)$ labeled by a sample number $i=1,2,3,\cdots$, the density matrix is given by
\begin{align}
\rho(\vec x,\vec y,t) = \lim_{N\to \infty}\frac{1}{N}\sum_{i=1}^N \psi_i(\vec x,t)\psi_i^*(\vec y,t).
\end{align}
In this way, one can avoid the huge numerical memory and cost needed to store and evolve the density matrix.
Instead, one needs to produce a large ensemble of wave functions to reduce the statistical errors.
Among the stochastic unravelling methods, two methods are widely used: Quantum State Diffusion (QSD) \cite{gisin1992quantum} and Quantum Jump \cite{Plenio:1997ep}.
The situation of numerical simulation is summarized in the Table \ref{tab:simulation}.

\begin{table}
\centering
\caption{Numerical simulations of Lindblad equation for quarkonium. The entry for ``Dissipation'' shows whether or not one includes $i\dot V_S/4T$ term in Eq.~\eqref{eq:Lindblad_op_qbm}.}
\label{tab:simulation} 
\begin{tabular}{c|c|c|l} 
Description & Setup & Dissipation & \qquad\qquad\qquad Numerical Method \\ \hline 
NRQCD & 1D, U(1) & no & Stochastic Potential \cite{Akamatsu:2011se, Kajimoto:2017rel}\\
              & 3D, U(1) & no & Stochastic Potential \cite{Rothkopf:2013kya}\\
              & 1D, SU(3) & no & Stochastic Potential \cite{Sharma:2019xum, Akamatsu:2021vsh}\\
              & 1D, U(1) & yes & Quantum State Diffusion \cite{Akamatsu:2018xim, Miura:2019ssi} \\
              & 1D, SU(3) & yes & Quantum State Diffusion (this work)\\
              & 1D, U(1) & yes & Direct evolution \cite{Alund:2020ctu}\\ \hline 
pNRQCD & 1$_{+}$D, SU(3) & no & Direct evolution for S and P waves \cite{Brambilla:2016wgg, Brambilla:2017zei}\\
                & 3D, SU(3) & no & Quantum Jump \cite{Brambilla:2020qwo,Brambilla:2021wkt}
\end{tabular}
\end{table}

Hereafter, we show the first result of the simulation of Lindblad equation from NRQCD description (with color SU(3) and including the quantum dissipation), which is solved by the QSD method.
In the QSD method, in order to simulate \eqref{eq:Lindblad}, one evolves the wave function stochastically by
\begin{align}
|\psi(t+dt)\rangle &= |\psi(t)\rangle
-i H_S'|\psi(t)\rangle dt
+\sum_k\left(
\langle L_k^{\dagger}\rangle_{\psi} L_k - \frac{1}{2}L_k^{\dagger}L_k 
- \frac{1}{2}\langle L_k^{\dagger}\rangle_{\psi}\langle L_k\rangle_{\psi}
\right) |\psi(t)\rangle dt \nonumber \\
& \quad +\frac{1}{\sqrt{2}}\sum_k \left(L_k - \langle L_k\rangle_{\psi}\right)|\psi(t)\rangle d\xi_k
\end{align}
with complex white noises $d\xi_k$ satisfying the following statistical properties
\begin{align}
\overline{d\xi_k}&=\overline{{\rm Re} (d\xi_k){\rm Im} (d\xi_{\ell})} = 0, \quad
\overline{{\rm Re} (d\xi_k){\rm Re} (d\xi_{\ell})}
=\overline{{\rm Im} (d\xi_k){\rm Im} (d\xi_{\ell})} = \delta_{k\ell} dt.
\end{align}
The quantum Brownian motion of quarkonium is modeled by the NRQCD Lindblad equation \eqref{eq:Lindblad_NRQCD} with $V(r)$ and $D(r)$ modeled by \eqref{eq:model}.
The simulation is performed on a one-dimensional lattice with $\Delta x=1/M$, $\Delta t=0.1M(\Delta x)^2$, and $N_x=254$ and the QGP temperature is $T=0.1M$.
The origin of the potential $C_FV(r)$ is singular in one dimension and is regulated by
\begin{align}
C_FV(r) = -\frac{0.3}{\sqrt{r^2+r_c^2}}e^{-2Tr}, \quad r_c = \frac{1}{M}.
\end{align}
As a reference time scale, let us quote the relaxation time $\tau_{\rm relax}$ of single heavy quark when the influence of QGP is modeled by $D(r)$:
\begin{align}
\frac{dp}{dt} = -\eta p, \quad
\eta = -\frac{C_F}{MT}\frac{d^2 D(x)}{dx^2}\Bigr|_0 = \frac{T^2}{\pi M} = \frac{M}{100\pi}, \quad
\tau_{\rm relax} = \frac{100\pi}{M}.
\end{align}

\subsection{Density matrix}
\begin{figure*}
\centering
\includegraphics[width=4cm,bb=60 0 320 216]{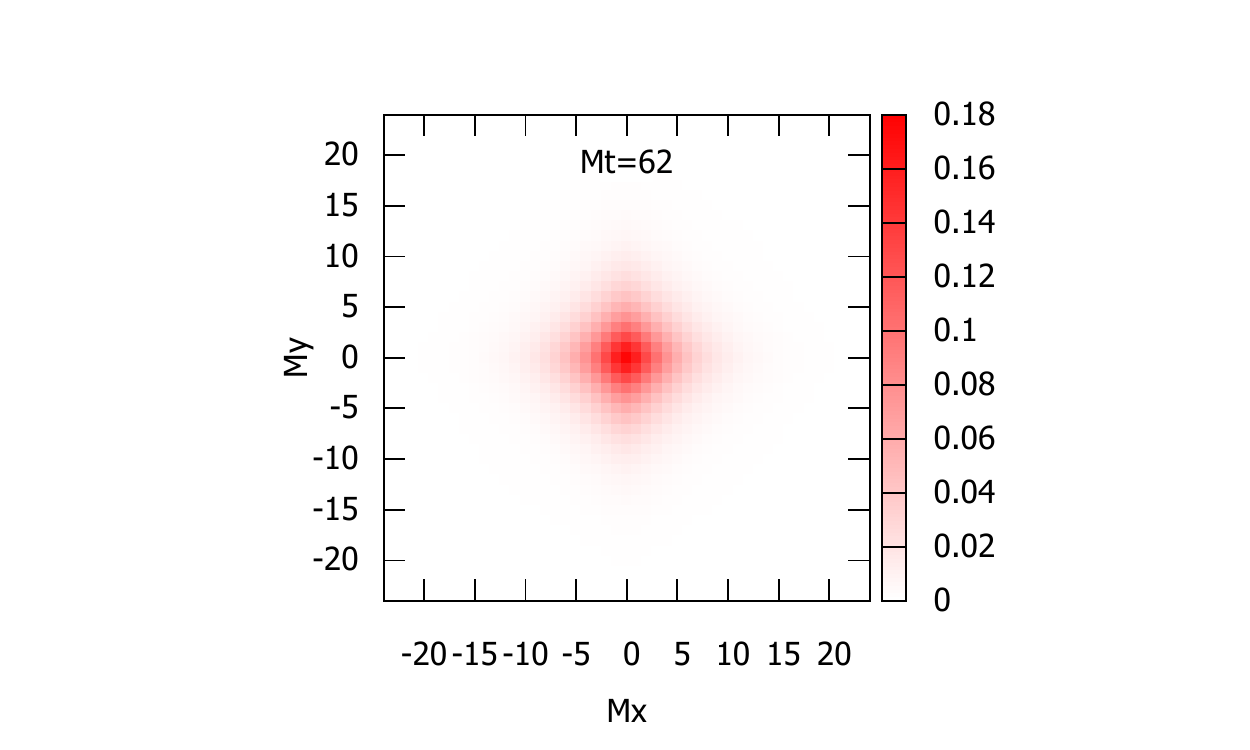}
\includegraphics[width=4cm,bb=60 0 320 216]{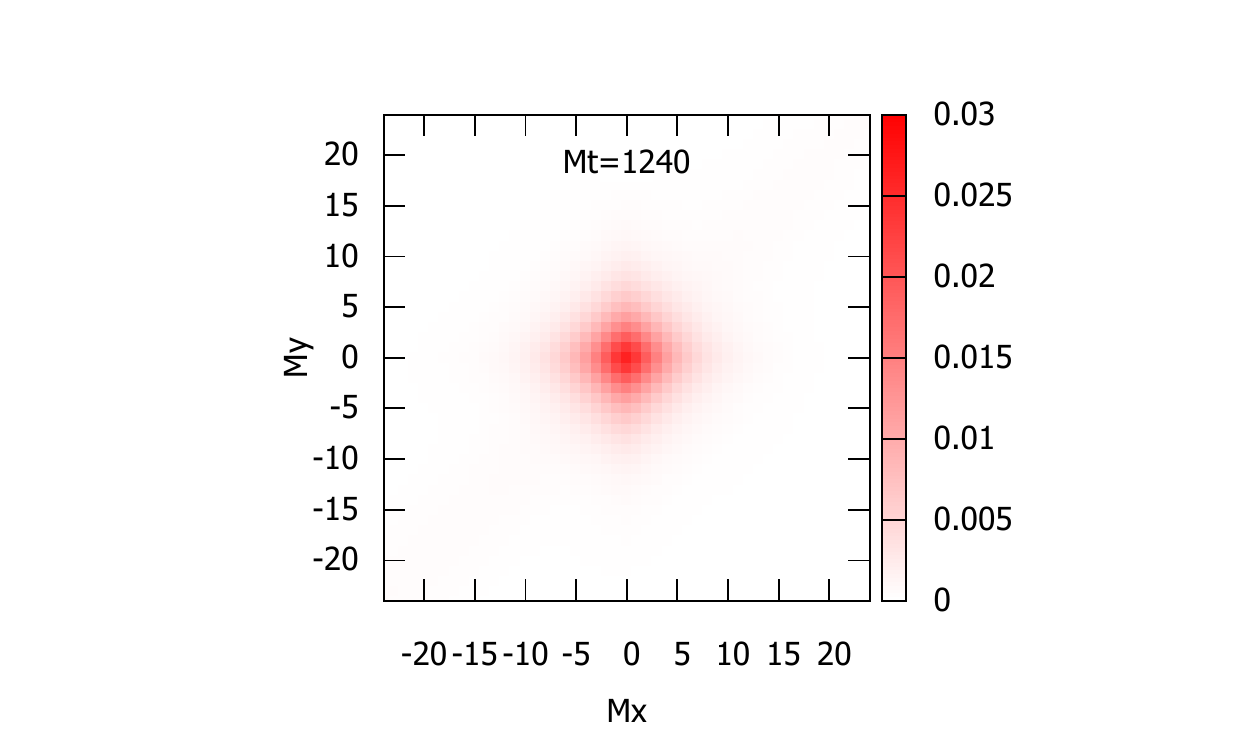}
\includegraphics[width=4cm,bb=60 0 320 216]{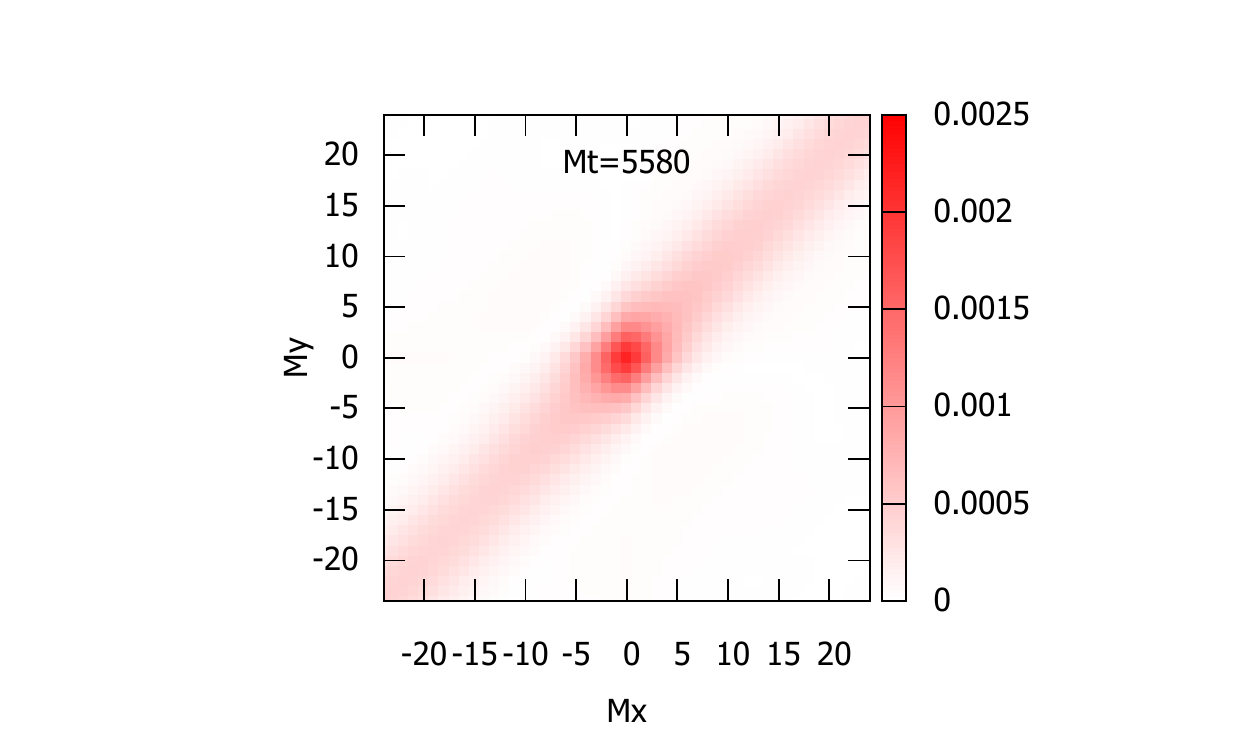}\\
\includegraphics[width=4cm,bb=60 0 320 216]{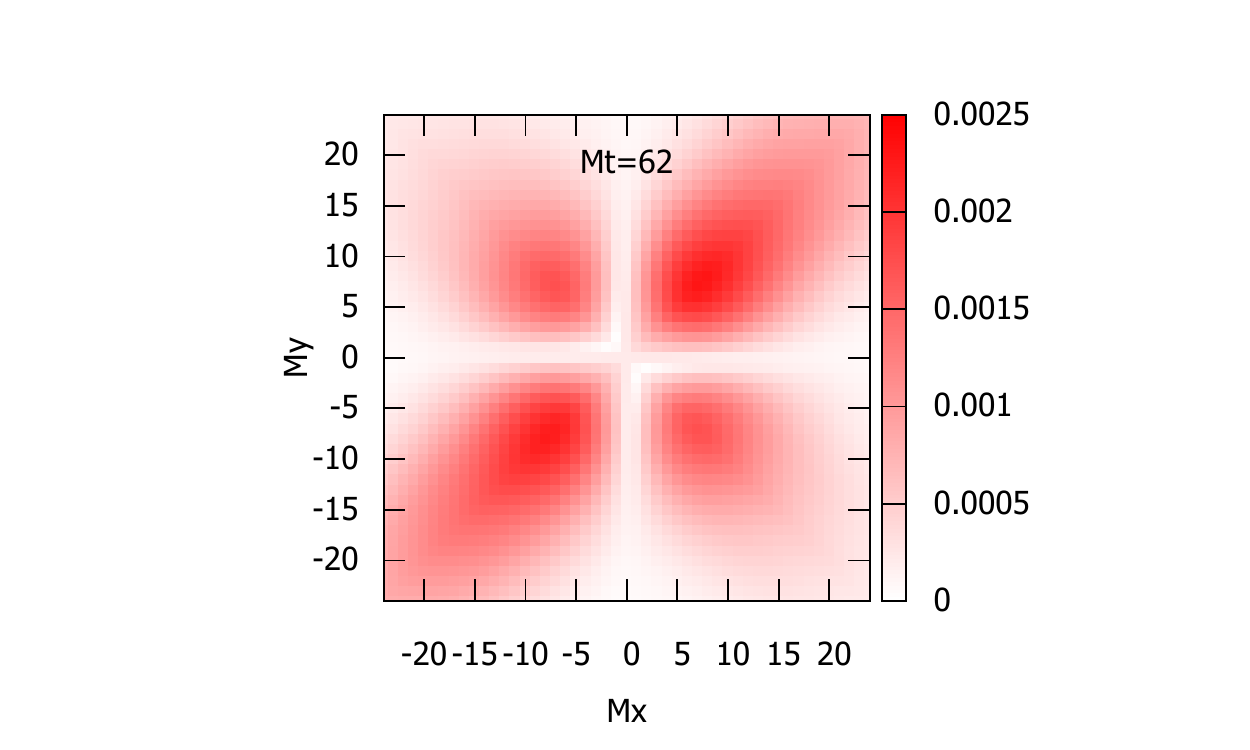}
\includegraphics[width=4cm,bb=60 0 320 216]{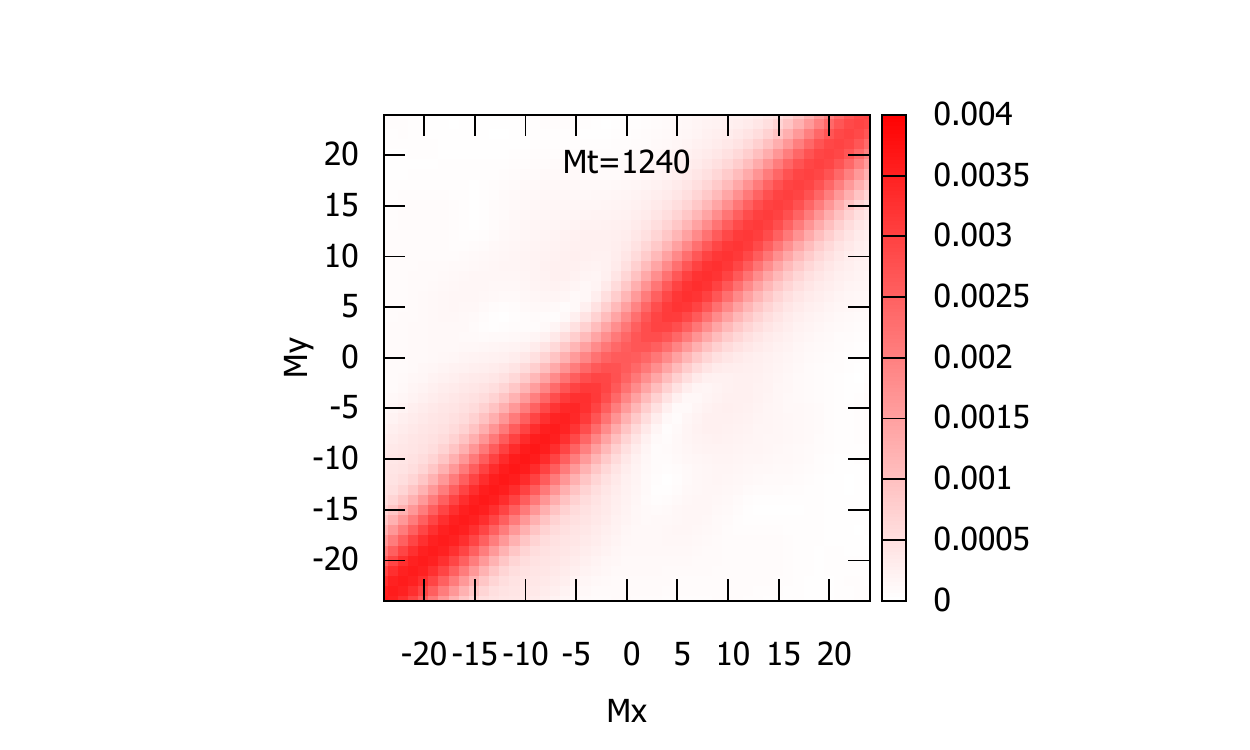}
\includegraphics[width=4cm,bb=60 0 320 216]{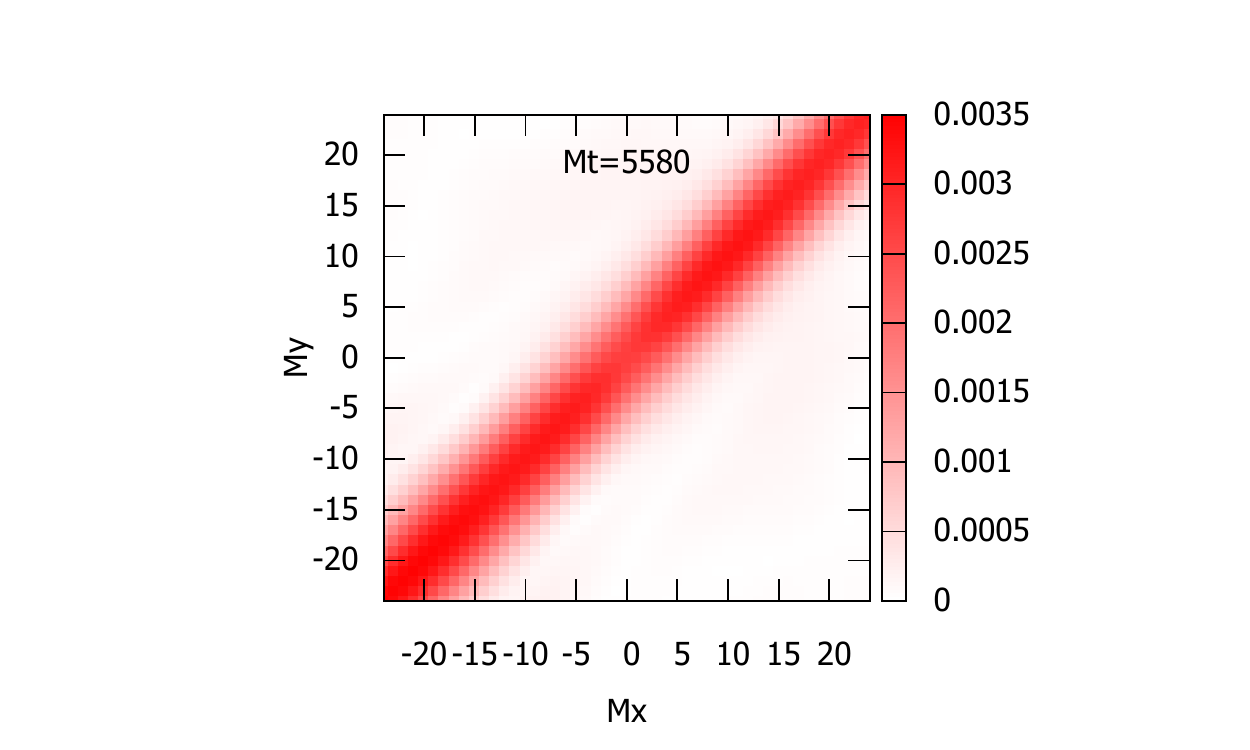}
\caption{Time evolution of the reduced density matrix in the singlet and octet sectors $|\rho_{s,o}(x,y,t)|^2/M^2$.
The figure only shows central domain $-24 \leq Mx, My \leq 24$.
The top figures are the single density matrices and the bottom figures are the octet density matrices.}
\label{fig:density_matrix}
\end{figure*}

In Fig.~\ref{fig:density_matrix}, time evolution of the density matrix is shown.
The top figures are the singlet density matrices and the bottom figures are the octet density matrices.
The initial wave function is the singlet ground state.
The evolution of the density matrix proceeds by the three steps.
\begin{enumerate}
\item[(1)] The singlet ground state is excited to octet as a color dipole (left).
\item[(2)] The octet density matrix is diagonalized (center).
\item[(3)] De-excitation from the octet to the singlet is observed and the system finally reaches a steady state (right).
\end{enumerate}

\subsection{Equilibration and the effect of dissipation}
\begin{figure*}
\centering
\includegraphics[width=6cm,bb=0 0 360 200]{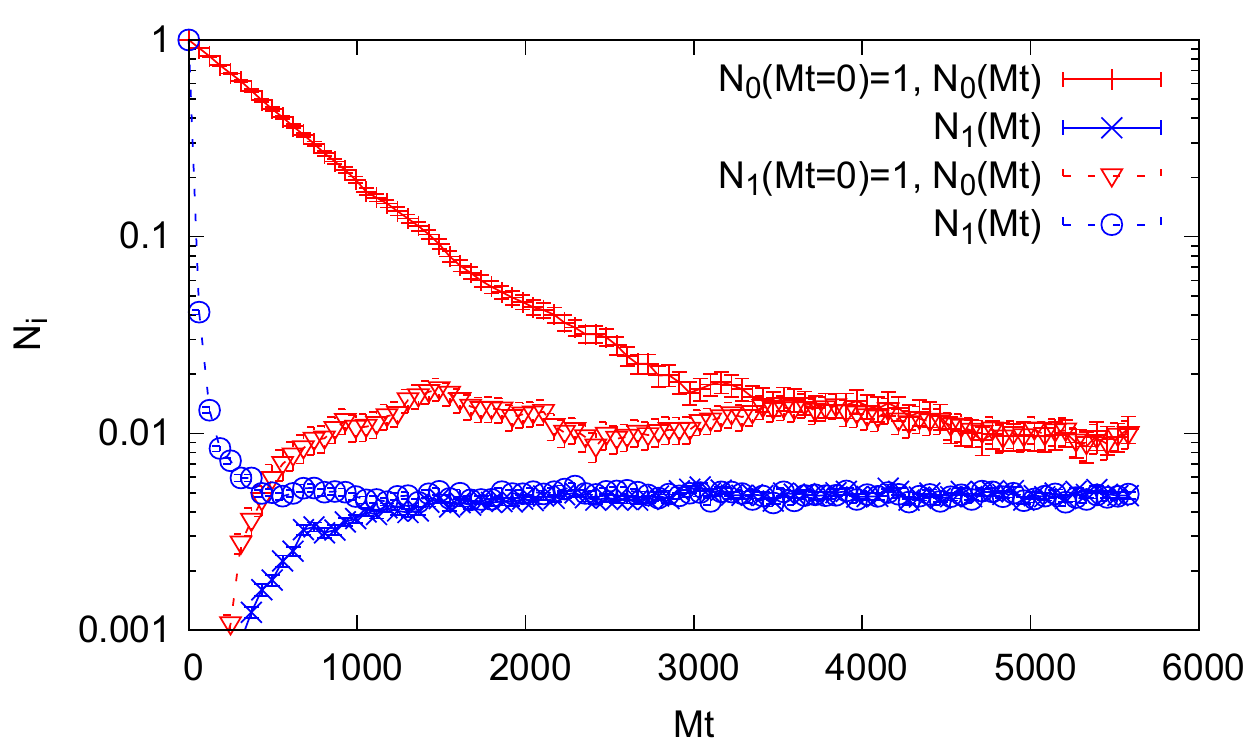}
\includegraphics[width=6cm,bb=0 15 340 180]{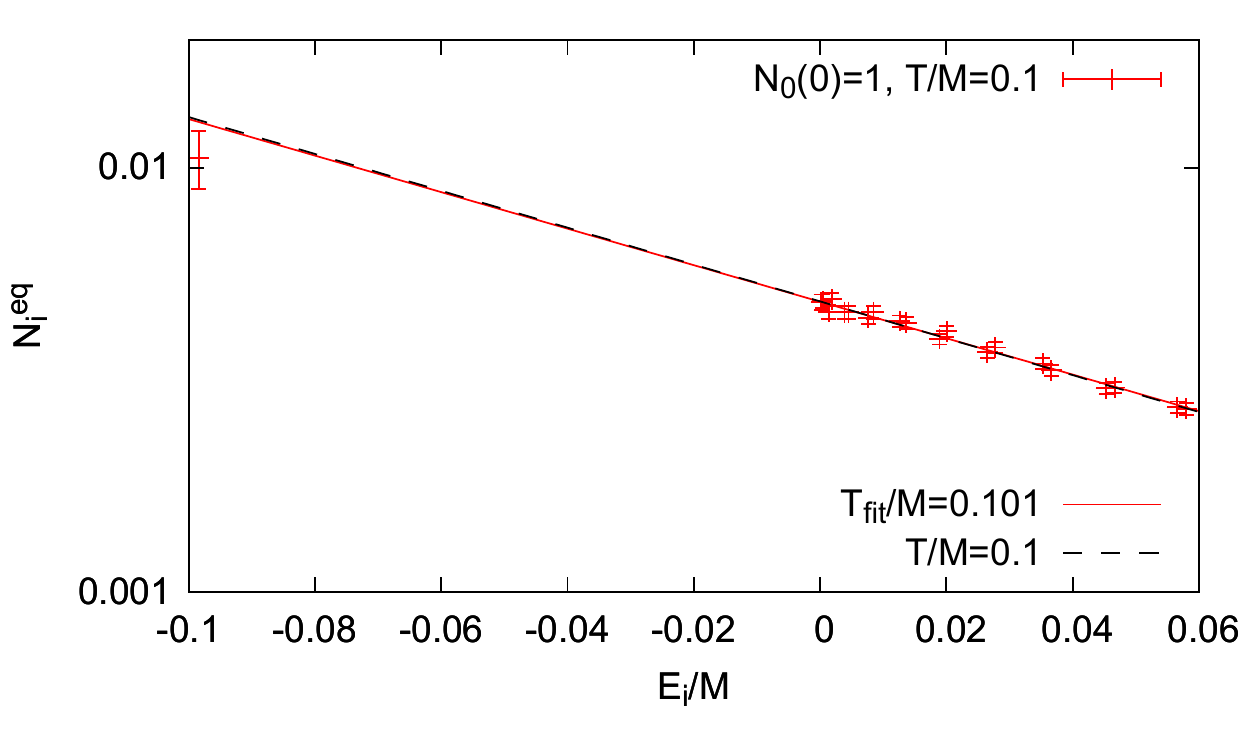}
\caption{Occupation probabilities for the ground state and the first excited state with different initial conditions (left).
Occupation of the lowest 20 levels (singlet) are compared with the Boltzmann distribution (right).}
\label{fig:equilibration}
\end{figure*}

The equilibration is not necessarily guaranteed in the Lindblad equation, at least it is not proven analytically.
Since the Lindblad equation is derived when QGP is in the equilibrium, it is natural to expect that the quarkonium system equilibrates.
In order to confirm the equilibration, let us show the occupation probabilities of the eigenstates.
In Fig.~\ref{fig:equilibration} (left), the occupation probabilities for the ground and the first excited states (both singlet) are plotted as a function of time.
Two initial conditions are employed, namely the ground state and the first excited state.
At around the last time step $Mt=5580$, the occupation does not depend on the initial conditions and reaches more or less a steady state.
In Fig.~\ref{fig:equilibration} (right), the occupation of the lowest 20 levels (singlet) is shown and is nicely fitted by the Boltzmann distribution with a temperature close to the environment $T_{\rm fit}=0.101M \simeq T$.

\begin{figure*}
\centering
\includegraphics[width=6cm,bb=0 0 360 200]{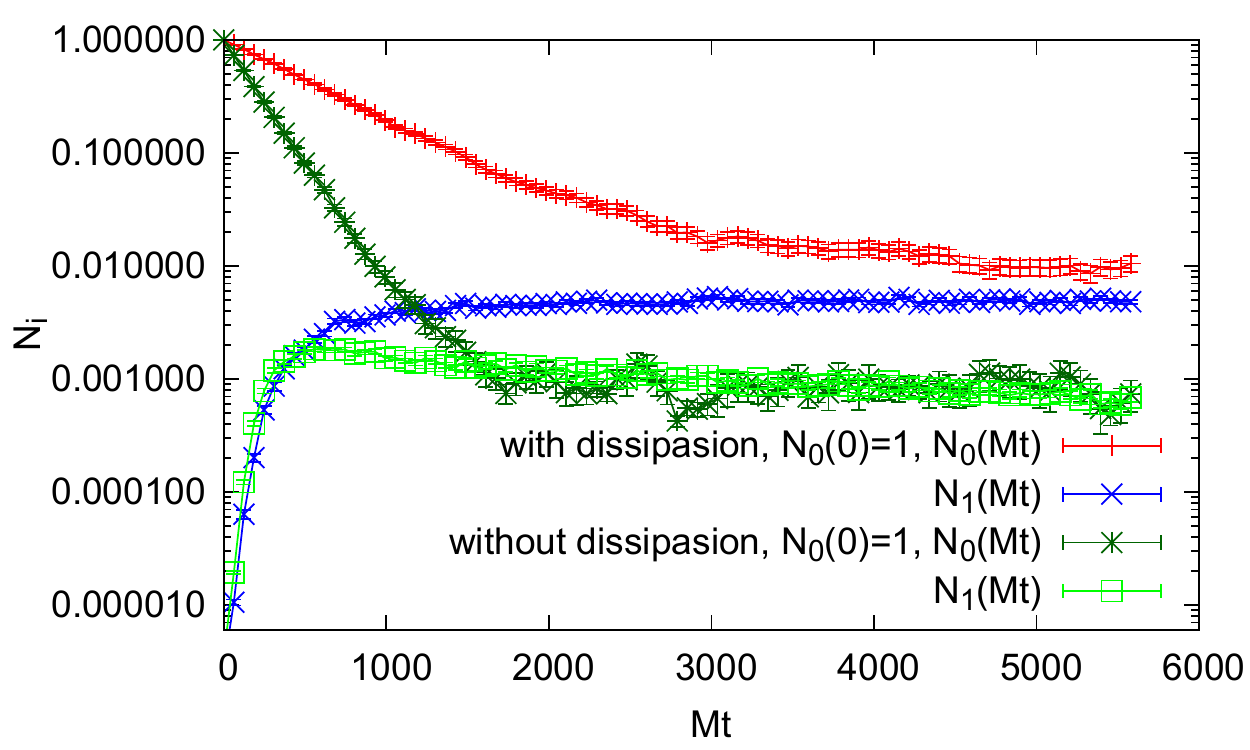}
\caption{Occupation probabilities for the ground state and the first excited state with and without the dissipative term, i.e. $i\dot V_S/4T$ term in Eq.~\eqref{eq:Lindblad_op_qbm}.}
\label{fig:wo_dissipation}
\end{figure*}

Finally, let us deliberately turn off the $i\dot V_S/4T$ term in Eq.~\eqref{eq:Lindblad_op_qbm}, namely simulate the Lindblad equation \eqref{eq:Lindblad_NRQCD} derived from NRQCD without $\mathcal O(\partial_t)$ corrections.
Figure \ref{fig:wo_dissipation} compares the results with and without this term when the initial condition is the singlet ground state.
As is mentioned in Sec.~\ref{sec:basics}, the term $i\dot V_S/4T$ can qualitatively change the dynamical property of the Lindblad equation.
From the Fig.~\ref{fig:wo_dissipation}, it is found that the long-time behavior does not show equilibration without this term and the deviation starts even at the very early time.
The lack of equilibration is simply because the requirement from the fluctuation-dissipation relation is not met.
The deviation at early time can be explained as follows.
For small wave function size, the decoherence, described by the first term in $L\propto V_S+i\dot V_S/4T$, does not proceed effectively.
In such a case, the relative importance of the second term is enhanced \cite{Akamatsu:2018xim, Miura:2019ssi}.
In the same way, the qualitative difference of the late behavior can also be understood.
Since the typical coherence length gets shorter and shorter by decoherence, at some stage the importance of the first and the second terms in $L$ balances and the system equilibrates.
Without the second term, the decoherence proceeds forever and the system never reaches proper equilibrium.

\section{Outlook}
\begin{itemize}
\item Applicability of the dipole approximation needs to be checked \cite{Miura-Kaida-et-al}.
Since the numerical cost of simulating the Lindblad equation from pNRQCD is much lower and since the Lindblad equation is beyond the weak coupling, it would be very fortunate if the dipole approximation is applicable, at least for finite but long enough time scale for heavy-ion phenomenology.
\item At lower temperature $T\sim T_c$, the applicability of the quantum Brownian regime to quarkonium dynamics is subtle.
It would be desirable to consider matching to another effective description by classical kinetic theory \cite{Yao:2017fuc, Yao:2018nmy, Yao:2018sgn, Yao:2020xzw}, which describes the quarkonium as a molecule with internal bound state levels.
\item Initial condition of quarkonium can be elaborated by e.g. simulating under the presence of thermalizing quarks and gluons and gauge fields.
\end{itemize}

\section*{Acknowledgments}
The work of Y.A. is supported by JSPS KAKENHI Grant Number JP18K13538.

%
 \bibliography{proceedings}
%
%
%
%

\end{document}